\documentclass[journal]{IEEEtran}

\usepackage{lineno} 

\usepackage{amsmath,amsfonts}
\usepackage{array}
\usepackage[caption=false,font=normalsize,labelfont=sf,textfont=sf]{subfig}
\usepackage{textcomp}
\usepackage{stfloats}
\usepackage{url}
\usepackage{verbatim}
\usepackage{graphicx}
\usepackage{cite}

\usepackage[bookmarksnumbered=true]{hyperref}
\hypersetup{
hidelinks,
colorlinks=true,
linkcolor=blue,
citecolor=blue,
urlcolor=black
}
\usepackage[linesnumbered,ruled,vlined]{algorithm2e}
\usepackage{bbm} 

\usepackage{amsthm} 

\usepackage{hyperref} 
\usepackage{xcolor}

\usepackage{enumerate} 

\begin{document}

\newcounter{probcounter}
\newcommand{\refporbcounter}[1]{\refstepcounter{probcounter}\theprobcounter \label{#1}} 

\title{Revolutionizing Wireless Networks with Self-Supervised Learning: A Pathway to Intelligent Communications
}

\author{
Zhixiang Yang, 
Hongyang Du,
Dusit Niyato,~\IEEEmembership{Fellow,~IEEE,}
Xudong Wang,
Yu Zhou,
Lei Feng,~\IEEEmembership{Member,~IEEE,} 
Fanqin Zhou,~\IEEEmembership{Member,~IEEE,}
Wenjing Li,~\IEEEmembership{Member,~IEEE,}
Xuesong Qiu,~\IEEEmembership{Senior Member,~IEEE}


\thanks{\textit{Corresponding author: Lei Feng.}}
\thanks{
Z. Yang, X. Wang, Y. Zhou, L. Feng, F. Zhou, W. Li and X. Qiu are with the State Key Laboratory of Networking and Switching Technology, Beijing University of Posts and Telecommunications, Beijing, China, 100876 
(e-mail: yangzx@bupt.edu.cn,  xdwang@bupt.edu.cn, zhouyu2020@bupt.edu.cn, fenglei@bupt.edu.cn, fqzhou2012@bupt.edu.cn, wjli@bupt.edu.cn, xsqiu@bupt.edu.cn).}
\thanks{Hongyang Du and Dusit Niyato are with the 
College of Computing and Data Science,
Nanyang Technological University, Singapore (e-mail: hongyang001@e.ntu.edu.sg; dniyato@ntu.edu.sg).}


}



\maketitle

\begin{abstract}

With the rapid proliferation of mobile devices and data, next-generation wireless communication systems face stringent requirements for ultra-low latency, ultra-high reliability, and massive connectivity. Traditional AI-driven wireless network designs, while promising, often suffer from limitations such as dependency on labeled data and poor generalization. To address these challenges, we present an integration of self-supervised learning (SSL) into wireless networks. SSL leverages large volumes of unlabeled data to train models, enhancing scalability, adaptability, and generalization. This paper offers a comprehensive overview of SSL, categorizing its application scenarios in wireless network optimization and presenting a case study on its impact on semantic communication. Our findings highlight the potentials of SSL to significantly improve wireless network performance without extensive labeled data, paving the way for more intelligent and efficient communication systems.

\end{abstract}


\section{Introduction}

With the explosive growth of mobile devices and data, various emerging vertical services, including smart cities and metaverse construction, pose more stringent requirements on next-generation wireless communication systems in terms of ultra-low latency, ultra-high reliability, and massive connectivity. For instance, metaverse services require six nines of connection reliability and sub-millisecond ultra-low interaction times \cite{liu2023slicing4meta}. 
In recent years, artificial intelligence (AI) has emerged as a highly promising paradigm in wireless network designs.
The powerful environmental adaptability and complex model learning capabilities of AI enable it to meet the diverse service requirements of various vertical industries.
Specifically, AI integration in wireless networks can automatically learn the characteristics of the entire wireless communication system and achieve end-to-end optimization by learning and modeling the complex relationships between variables \cite{du2023age}.

However, existing AI has certain shortcomings and limitations in several aspects. Most AI technologies, such as deep learning, often heavily rely on a large amount of labeled data samples, which can consume considerable time and costs, e.g., for labelling, in practical applications. Additionally, the current supervised learning architectures integrating AI into wireless networks are often designed for specific tasks and system scenarios, leading to issues with insufficient generalization capability. If task objectives or network scenarios change, existing AI algorithms need to be redesigned and retrained to meet new requirements \cite{zhou2022knowledge}. Moreover, emerging cross-modal communication applications, such as metaverse services integrate visual, auditory, and tactile signals to provide immersive interactive experiences for individuals. Due to the semantic differences and temporal asynchrony of multimodal data, existing AI technologies still face significant challenges in handling multimodal data.

To address the aforementioned issues, we propose a method of integrating self-supervised learning (SSL) into wireless networks. SSL selects its own supervisory information from a lot of unlabeled data and utilizes this information to train the deep neural network, thereby learning representations that are effective for various downstream tasks \cite{salihu2024self, baevski2022data2vec}. Although SSL has been introduced before, it has only recently garnered significant attention. Due to its strong generalization ability and efficient data processing capability, SSL has been widely applied in computer vision (CV), graph learning, and natural language processing.
SSL demonstrates potential in several aspects:
\begin{itemize}
    \item \textbf{Reduction in Dependence on Labeled Data:} SSL significantly reduces the need for  labeled data by leveraging the vast amounts of unlabeled data available. This makes SSL more cost-effective and scalable, as acquiring labeled data can be expensive, time-consuming or even impossible in practical applications \cite{jing2021selfsupervised}.

    \item \textbf{Improved Generalization Capability:} SSL models learn from diverse and extensive datasets, capturing a wide range of patterns and variations. This leads to the development of robust representations that generalize well to new and unseen data, enhancing the overall performance of the SSL model in real-world applications. SSL has been widely applied in foundational models for generative AI such as large language models (LLMs), generative adversarial networks (GANs), and variational auto-encoders (VAEs) \cite{jenni2018self}.

    \item \textbf{Transferability of Learned Representations:} The representations learned through SSL are highly transferable to various downstream tasks. This means that a model pre-trained using SSL can be fine-tuned on specific tasks with minimal additional labeled data, providing strong accuracy across different applications and reducing the need for extensive task-specific training \cite{liu2021self}. 
    
\end{itemize}

In this paper, we integrate SSL into wireless networks to achieve communication intelligence. To the best of our knowledge, this is the first work to discuss a comprehensive review, case study, and future directions of SSL in wireless networks. Our contributions can be summarized as follows:

\begin{itemize}
    \item We present a comprehensive overview of SSL, including the importance of SSL, and an investigation of its applications in CV and LLM. Then, we discuss the reasons why SSL is specially suitable for wireless applications.
    \item We classify and summarize application scenarios of SSL in wireless communications and networking, including channel estimation and prediction, anomaly detection, signal classification, network optimization, user mobility behavior prediction, interference management, etc.
    We also present potential wireless applications.
    \item To further explore the benefits of SSL in wireless network, we conduct a case study that presents advantages of SSL on semantic communication.
    Specifically, we use the generative SSL to enhance the accuracy of semantic communication information reconstruction without requiring large amounts of labeled data.
\end{itemize}

\section{Overview of Self-Supervised Learning}

In this section, we elaborate the basic principles of SSL and its application in  CV, LLMs and wireless networks.

\subsection{Basics of Self-Supervised Learning}




Machine learning has made great strides and advancements. As two pivotal methods within machine learning, supervised learning and unsupervised learning have been widely applied in numerous fields. The main difference between these two lies in whether the input data is labeled. 
Supervised learning relies heavily on manual labeling. Thus, it is usually trained to ``learn" the mapping relationship between inputs and outputs through the labeled data to make predictions or classifications based on unobserved data. In contrast, there is no need for label in unsupervised learning. It employs clustering techniques that is able to uncover latent structures and hidden patterns within the data to overcome the dependency on label \cite{gupta2022overview}.

As one of the most extensively studied areas in machine learning, supervised learning is currently encountering daunting challenges. On the one hand, it suffers from several drawbacks caused by manual labeling, such as low efficiency and high costs. For instance, Scale.ai, a company specializing in data labeling services, charges  \$6.4 per image for labeling in 2020 \footnote{Data Sources: 
https://medium.com/whattolabel/data-labeling-ais-human-bottleneck-24bd10136e52}, which implies that the cost of labeling a high-quality image dataset could easily reach hundreds of thousands or even millions of dollars. On the other hand, a series of other issues, such as generalization error, spurious correlations and adversarial attacks, are prevalent, and it is significantly complicated to resolve them \cite{liu2021selfsupervised}. In contrast, while unsupervised learning is not reliant on labeled data, it faces its own set of difficulties due to the lack of prior knowledge. Determining an optimal number of clusters during the process is an especially challenging problem for unsupervised learning. Furthermore, there are additional challenges including representation of data objects, establishment of evaluation criteria, and extraction of knowledge \cite{ezugwu2022comprehensive}. These factors constitute the main reasons why the research and application of unsupervised learning proceed at a relatively slower pace.

\begin{figure*}
    \centering
    \includegraphics[width=0.95\linewidth]{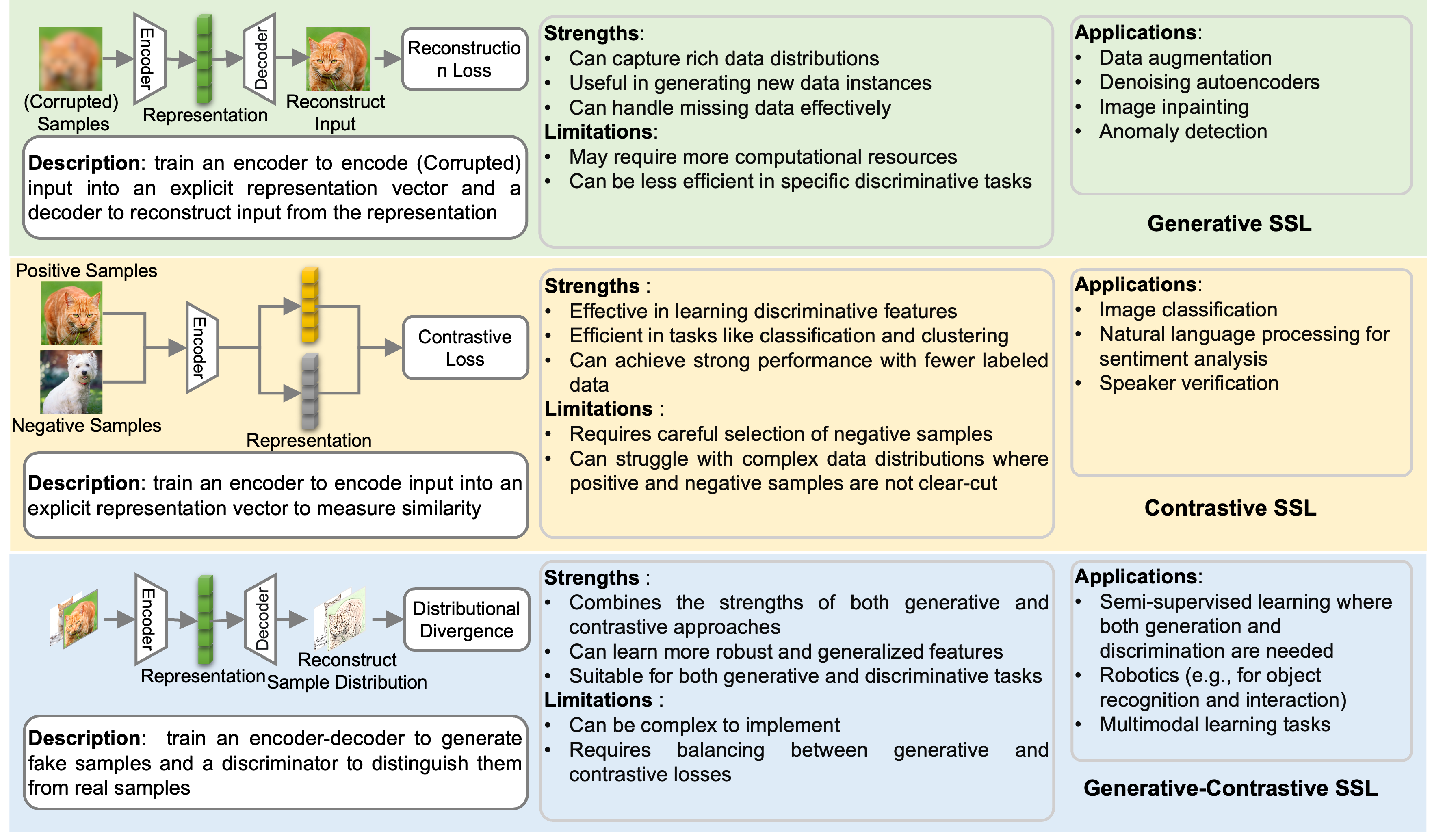}
    \caption{Comparison of Self-Supervised Learning Methods: Generative, Contrastive, and Generative-Contrastive SSL Approaches. This analysis provides an in-depth examination of their mechanisms, advantages, disadvantages, and applications. Generative SSL is primarily designed for input reconstruction, making it ideal for tasks like data augmentation. Contrastive SSL excels in classification tasks by learning discriminative features. Generative-Contrastive SSL combines both approaches, offering robust solutions for multimodal learning tasks and achieving a balance between generative and discriminative objectives.}
    \label{fig_categories_ssl}
\end{figure*}

Given the inherent limitations of supervised and unsupervised learning, SSL has attracted much attention from researchers recently. The basic principle of SSL is similar to supervised learning in that the model is also trained through labeled data to accomplish a supervised downstream task. However, what sets SSL apart is that the data is labeled by itself rather than humans. SSL mainly consists of two components: pretext task and downstream task. In the pretext task phase, the model learns useful representations akin to how it would in supervised learning, but its labels based on the intrinsic properties of the data. To ensure the quality of the downstream task, the model typically fine-tunes the representations learned from the pretext task according to the specific requirements of the downstream task, thereby completing the supervised downstream task effectively. In fact, SSL proves highly effective for complex tasks such as image classification, object detection, and image segmentation, along with tasks having limited labels. Creating an efficient and meaningful pretext task is often regarded as a decisive factor influencing the performance of SSL \cite{rani2023selfsupervised}.
According to the objectives of SSL, mainstream SSL methods can be divided into three categories with their main structural differences illustrated in Fig. \ref{fig_categories_ssl} \cite{liu2021selfsupervised}:
\begin{itemize}
   \item \textbf{Generative SSL}: 
   Generative SSL can enable models to capture essential features and latent structures within data through reconstruction or prediction tasks to obtain useful representations that are applicable to a variety of downstream tasks.
   \item \textbf{Contrastive SSL}: 
   Contrastive SSL are adept at learning relative positions or affiliation relations between local and global aspects, finding broad applications in fields such as speech recognition and image classification. 
   \item \textbf{Generative-Contrastive (Adversarial) SSL}: 
   Adversarial SSL learns to reconstruct the original data distribution by minimizing distribution divergence rather than individual samples. The development of Adversarial SSL is inseparable from Generative Adversarial Networks (GAN). Adversarial SSL finds applications in multiple domains such as capturing full information of samples, recovering content from partial inputs, pre-training language models, and graph learning.
\end{itemize}

\subsection{Self-Supervised Learning for Computer Vision}

\subsubsection{The importance of SSL in CV}


In the field of CV, SSL leverages vast amounts of unlabeled image data to learn rich feature representations. 
These are crucial for downstream tasks such as image classification and object detection, enabling effective model training without extensive labeled datasets.

\begin{itemize}
    \item \textbf{Cost and Scalability of Data Annotation:} 
    SSL reduces reliance on manual labeling by leveraging unlabeled data to learn useful feature representations. 
    This approach significantly lowers the cost of data labeling and enhances data utilization by enabling models to learn from large-scale unlabeled datasets.

    \item \textbf{Generalization Capability of Models:} SSL improves model generalization by learning the intrinsic structures and patterns of data. 
    For instance, by predicting the missing content of an image or rearranging shuffled image segments, the features learned by the model are more robust and generalize better to new data set.
\end{itemize}


\subsubsection{Application of SSL in CV}
Models such as SimCLR and MoCo learn by contrasting positive pairs (similar or identical images) against negative pairs (different images). 
SimCLR \cite{chen2020simple} demonstrates the effectiveness of feature learning through contrastive loss on large datasets.
MoCo \cite{he2020momentum} effectively improves performance across various visual tasks, including image classification and object detection, by constructing a dynamic dictionary for contrastive learning.
This method effectively teaches the model to understand which features are critical for distinguishing between different images, greatly improving the efficiency and quality of the learned representations.

\subsection{Self-Supervised Learning for Large Language Models}
\subsubsection{The importance of SSL in LLMs}
In the field of LLMs, SSL enables models to learn from vast amounts of textual data without requiring explicit annotations, which is crucial given the expansive and continually growing datasets used today. 

\begin{itemize}
    \item \textbf{Efficient Use of Unlabeled Data:} SSL allows for the effective use of this vast reservoir of unlabeled raw text, enabling models to learn from the inherent structure and context of language without the need for extensive manual annotation.
    \item \textbf{Understanding of Language Nuances:} SSL techniques enable models to learn the context and semantics of words and phrases from their co-occurrence and placement within large text corpora. For example, models learn to predict missing words or the next word in sentences, gaining a deeper understanding of language structures and nuances.
\end{itemize}

\subsubsection{Application of SSL in LLMs}
In the realm of natural language processing, SSL has revolutionized the development of large language models. 
Models like GPT (Generative Pre-trained Transformer) and BERT (Bidirectional Encoder Representations from Transformers) utilize vast amounts of unlabeled text to learn language representations. 
GPT, for example, uses a generative approach where it predicts the next word in a sequence, learning contextual relationships between words. BERT, on the other hand, uses a masked language model (MLM) approach where some words in a sentence are randomly masked, and the model learns to predict the masked word based on its context.

This self-supervised training enables these models to develop a deep understanding of language syntax, semantics, and even some aspects of common-sense reasoning, which can then be fine-tuned for specific tasks such as sentiment analysis, question answering, and more.

\subsection{Workflow for SSL}

\begin{figure*}
    \centering
    \includegraphics[width=0.85\linewidth]{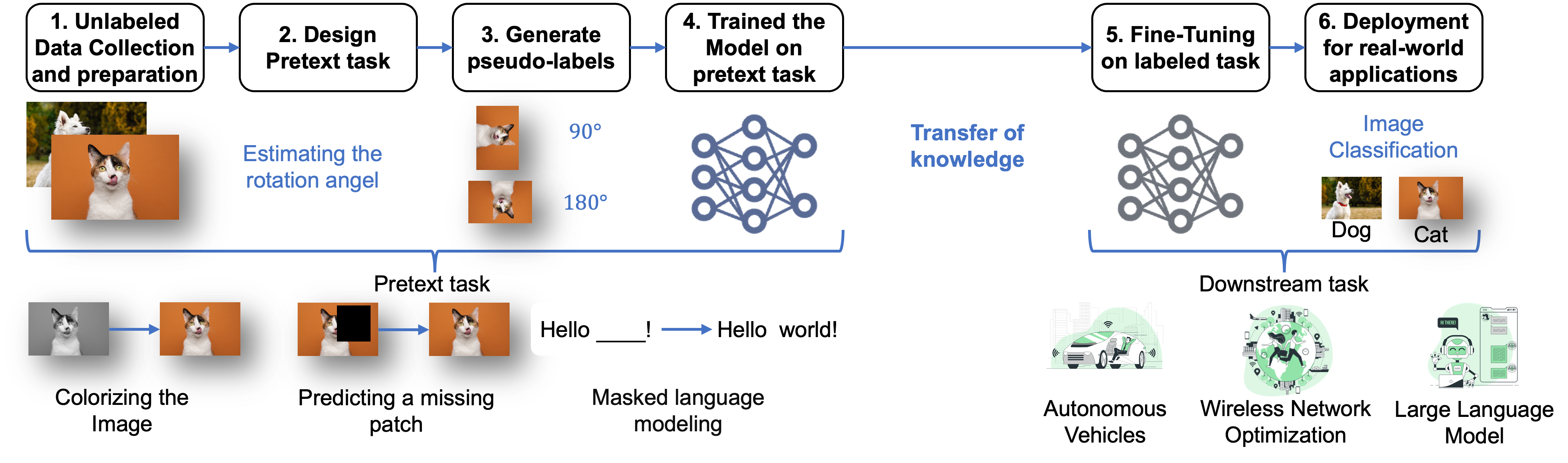}
    \caption{The workflow diagram for SSL illustrating the process where an appropriate pretext task is employed to pre-train the model. This pre-trained knowledge is then transferred and refined through fine-tuning on labeled data, enabling the application of SSL to a variety of downstream tasks.}
    \label{fig_ssl_tutorial}
\end{figure*}

SSL leverages unlabeled data to train models through several key steps,  making it particularly valuable for wireless communication systems, which often generate large volumes of data that are not readily labeled. The structured workflow shown in Fig. \ref{fig_ssl_tutorial} elucidates the SSL process:

\textbf{1. Data Collection and Preparation}
\begin{itemize}
    \item Collection: 
    The first step involves acquiring a large dataset of unlabeled data pertinent to the target domain. In the context of wireless communication, this could include network traffic data, signal strengths, channel state information (CSI), or user behavior patterns.
    \item Preprocessing: 
    Before feeding the data into the model, it needs to be prepared appropriately. This involves:
    1) Normalization: Adjusting the data to a common scale without distorting differences in the ranges of values.
    2) Tokenization: For text data, breaking down the text into smaller units such as words or subwords.
    3) Resizing: For image data, adjusting the size of images to a uniform dimension suitable for the model.
\end{itemize}

\textbf{2. Design Pretext Task: }
Designing a pretext task is crucial in SSL. A pretext task exploits the inherent structure of the data to generate pseudo-labels, which the model uses to learn useful features. Examples include:
\begin{itemize}
    \item Image Data: 
    Tasks like predicting the rotation angle of an image, filling in missing parts (inpainting), or converting grayscale images to color.
    \item Text Data: 
    Tasks such as masked language modeling, where certain words in a sentence are masked and the model predicts them, or next sentence prediction, where the model predicts whether two sentences are sequential.
\end{itemize}

\textbf{3.Generate Pseudo-Labels:} Based on the pretext task, pseudo-labels are created. These labels are not manually annotated but are derived from the data itself. For instance:
\begin{itemize}
    \item In an image rotation task, the pseudo-labels could be the angles $(0^\circ, 90^\circ, 180^\circ, 270^\circ)$.
    \item In a masked language modeling task, the pseudo-labels are the words that the model needs to predict.
\end{itemize}


\textbf{4. Train the Model on Pretext Task}
\begin{itemize}
    \item Training: The model is trained using the pseudo-labeled data, focusing on solving the pretext task. This phase helps the model to learn useful representations and features from the data.
    \item Optimization: Appropriate loss functions and training algorithms are applied to optimize the model. For example, cross-entropy loss might be used for classification tasks, while mean squared error could be used for regression tasks.
\end{itemize}


 
\textbf{5.Fine-Tuning and Optimization: }After the model has learned from the pretext task, it undergoes fine-tuning.
\begin{itemize}
    \item Refinement: The model is fine-tuned on a smaller, labeled dataset to improve its performance on the actual downstream task, such as classification or prediction in wireless communication.
    \item Hyperparameter Tuning: Hyperparameters, such as learning rate, batch size, and the number of layers, are adjusted to further optimize the performance of the model.
\end{itemize}
	
\textbf{6. Deployment: }Finally, the trained and fine-tuned model is deployed in real-world applications.
\begin{itemize}
    \item Deployment: Implementing the model in practical scenarios, such as network traffic management, signal processing, or user behavior prediction in wireless networks.
    \item Monitoring and Updating: Continuously monitoring the performance of the model to ensure it meets the desired criteria and updating it with new data as required to maintain or improve performance.
\end{itemize}

This workflow enables SSL to harness the vast amounts of unlabeled data available, learning robust and informative representations that significantly enhance the performance of machine learning models across various applications.

\subsection{Why SSL is good for wireless and potential applications}


\subsubsection{Challenges in Wireless Networks} 
We summarize the current major challenges of machine learning in wireless networks as follows.

\begin{itemize}
    \item \textbf{Lack of Labeled Data}: 
    In many real-world scenarios, the data generated by wireless devices are not accompanied by labels.
    For example, it is difficult to obtain the ground truth of wireless channels in the real world.
    Network anomalies by their nature are rare and not well-defined.
    
    
    \item \textbf{Dynamic Environments}:  
    Wireless environments are highly dynamic, with changing network conditions, mobility, and interference patterns, which makes traditional supervised learning and other methods have to train and update models more frequently to ensure accuracy.
    For example, the presence of noise, interference, and signal distortion can significantly degrade classification performance.


    \item \textbf{Scalability and Deployment}: 
    Deploying machine learning models across diverse devices can be challenging due to the variability in data characteristics.
    For example, the existing anomaly detection algorithms cannot mine specific features for different abnormal types, which results in the weak generalization ability.


\end{itemize}

\subsubsection{Advantages of SSL in Wireless Communications} SSL holds considerable promise for the field of wireless communications, where data is abundant and unlabeled.
Here are key advantages of using SSL in wireless communications:
\begin{itemize}
    \item \textbf{Cost-Effectiveness}: 
    SSL can address the issue of lacking labeled data and achieve cost-effectiveness by automatically generating pseudo-labels from the vast amount of unlabeled data generated by user interactions, signal traffic, and network operations.

    \item \textbf{High Scalability and Adaptability}:
    SSL can adapt to the dynamic changes and overcome the challenges of dynamic environments by continuously learning from the ongoing data stream.
    


    \item \textbf{Generalization}: 
    Due to its powerful ability to generate, models trained with SSL tend to generalize better to new devices and unseen scenarios, thereby enhancing the generalization and deployment capabilities.

\end{itemize}

In summary, the ability of SSL to learn from unlabeled data, adapt to dynamic environments, and discover complex feature interactions makes it particularly suitable for tackling a wide range of problems in wireless communications.


\section{Self-Supervised Learning and Wireless Networks}

In this section, we summarize the current research of SSL in wireless network and analyze potential applications.

\subsection{Current Research of SSL in Wireless Network}

\begin{figure*}
    \centering
    \includegraphics[width=1.0\linewidth]{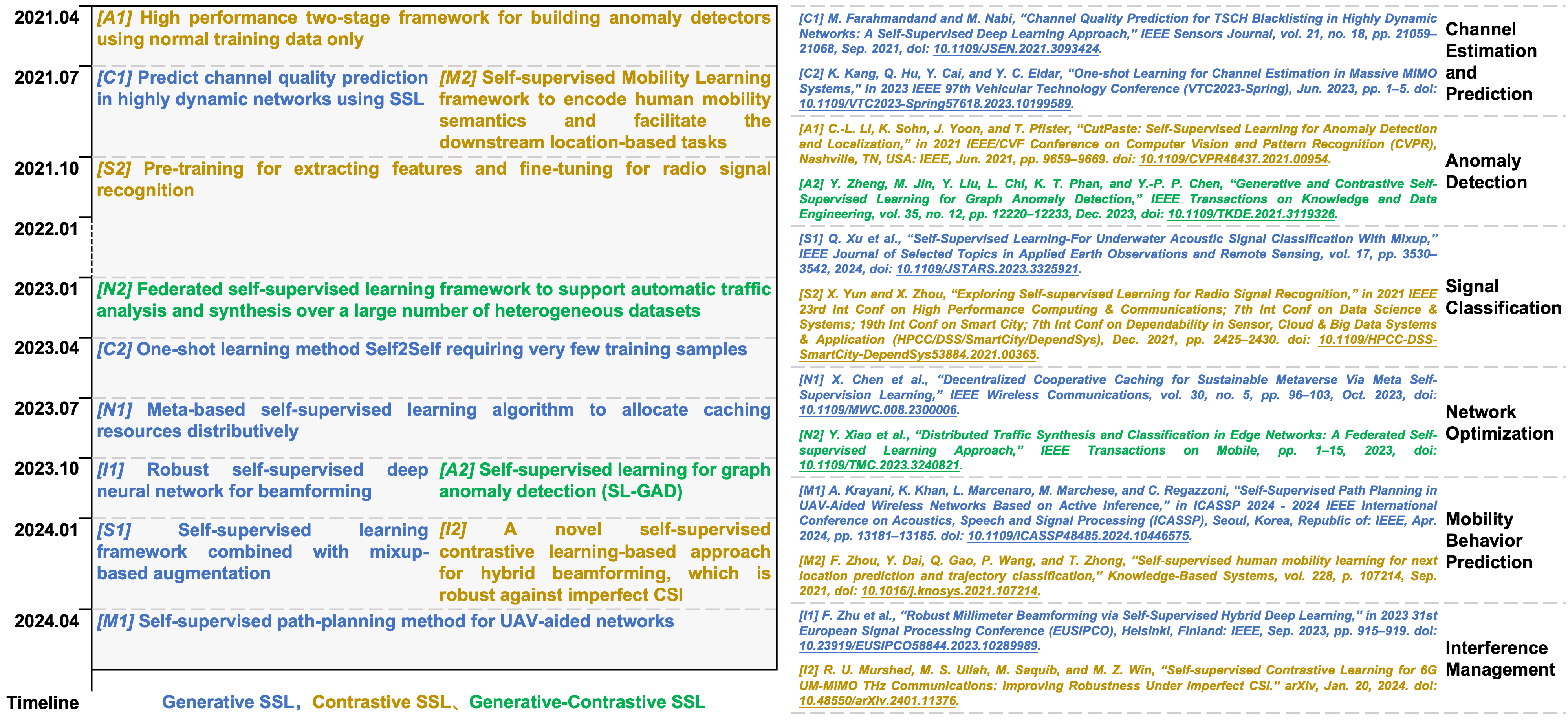}
    \caption{Summary of research topics and key contributions in wireless network optimization, where the blue, yellow and green fonts correspond to generative SSL, contrastive SSL and generative-contrastive SSL methods, respectively.}
    \label{fig_summary_table}
\end{figure*}

The Fig. \ref{fig_summary_table} provides a summary of research topics and key contributions in wireless network optimization and corresponding SSL methods.

\subsubsection{Channel Estimation and Prediction}
The task involves crucial techniques used to understand and anticipate the behavior of the communication channel through which data is transmitted.
The study in \textbf{\textit{[C1]}} of Fig. \ref{fig_summary_table} introduces a self-supervised method for training deep neural networks to predict future frequency channel behavior, performing similarly to enhanced time-slotted channel hopping in low-interference environments.
A one-shot SSL framework consisted of a traditional channel estimation module and a denoising module for channel estimation is proposed in \textbf{\textit{[C2]}} , which achieves very similar performance to the supervised learning approach.

\subsubsection{Anomaly Detection}
This task involves identifying patterns or occurrences in the network behavior that deviate from what is considered normal. 
The study in \textbf{\textit{[A1]}} 
develops a high-performance defect detection model that identifies unknown anomalies in images without anomalous data. Using the CutPaste data augmentation strategy to classify normal data and transfer learning on pretrained ImageNet representations, the model achieves a new state-of-the-art (SOTA) AUC of 96.6.
A novel SSL method for graph anomaly detection is proposed in \textbf{\textit{[A2]}} , which 
constructs various contextual subgraphs based on a target node and uses generative attribute regression and multi-view contrastive learning for anomaly detection. Experiments on six benchmark datasets show that the proposed method significantly outperforms SOTA methods (AMEN, CoLA, DOMINANT, ANOMALOUS, Radar, DGI).

\subsubsection{Signal Classification}
The task involves the identification and categorization of various types of signals transmitted over the network.
The study in \textbf{\textit{[S1]}} 
proposes a novel SSL framework for underwater acoustic signal classification, incorporating mixup-based augmentation to learn discriminative representations from large-scale unlabeled data, reducing the reliance on labeled data. This approach achieves a classification accuracy of $86.33~\%$ on the DeepShip dataset, significantly outperforming previous competitive methods.
The study in \textbf{\textit{[S2]}} 
The research introduces a two-step self-supervised representation learning framework using vast amounts of unlabeled data. The first step involves pre-training for feature extraction, while the second step fine-tunes the model for radio signal recognition. This model surpasses a supervised learning model trained on $10~\%$ labeled data, achieving a $9.3~\%$ higher accuracy rate, particularly at an SNR of $39.0~$dB.


\subsubsection{Network Optimization}
This task refers to the process of adjusting the configuration and operation of the network to maximize overall performance, efficiency, and quality of service.
The study in \textbf{\textit{[N1]}} 
explores an online distributed caching model within the digital twin metaverse and introduces a meta-based SSL algorithm. This algorithm enables a distributed allocation of caching resources across multiple small base stations, also known as metaverse edge nodes, to efficiently handle the high demand for ultra-low latency services.
The study in \textbf{\textit{[N2]}} 
introduces FS-GAN, a federated self-supervised learning framework, designed to automate traffic analysis and synthesis across a vast amount of heterogeneous datasets. Simulation results highlight that FS-GAN achieves an average classification accuracy improvement of over $20~\%$ compared to SOTA clustering-based traffic analysis algorithms. Furthermore, the proposed method exhibits a robust reduction in computation time by at least $70~\%$, while maintaining minimal loss in solution quality.


\subsubsection{User Mobility Behavior Prediction}
This task involves forecasting the movements and location patterns of users within the network.
The study in \textbf{\textit{[M1]}} 
proposes a universal self-supervised path-planning approach for unmanned aerial vehicle (UAV)-assisted networks. This method enables the UAV to anticipate the consequences of its actions through a world model and evaluate anticipated surprises in a self-supervised fashion.
The study in \textbf{\textit{[M2]}} 
proposes a self-supervised framework for learning human mobility patterns. This framework encodes mobility semantics to support location-based applications. It focuses on modeling noisy and sparse mobility trajectories, leveraging spatio-temporal data and augmented traces to refine trajectory representations.

\subsubsection{Interference Management}
This task is to mitigate the impact of interference on communication quality and network performance. 
The study in \textbf{\textit{[I1]}} 
The study proposes a self-supervised deep neural network for beamforming, which is rigorously tested in two datasets (new WAIR-D and DeepMIMO) representing different wireless deployment environments. Simulation results show that the network, utilizing hybrid learning, performs well in both datasets, demonstrating robust performance across varying environments.
The study in \textbf{\textit{[I2]}} 
The study introduces a novel self-supervised contrastive learning approach for hybrid beamforming that is resilient to imperfect CSI. Particularly, at a CSI SNR of $20~\mathrm{dB}$, the proposed method maintains stable spectral efficiency (SE), significantly outperforming traditional alternating minimization beamforming algorithms that exhibit a substantial seventeen-fold drop in SE.





\subsection{Potential applications}


\subsubsection{Semantic Communications}
In semantic communications, the focus is on transmitting the meaning or semantics of the message rather than the message itself. SSL can be used to learn semantic representations directly from the data, improving communication efficiency and effectiveness, especially in environments with limited bandwidth or noisy channels.
By learning to extract and transmit only the essential information, SSL can reduce the data needed for effective communication and enhance understanding between different systems or agents \cite{du2023semantic}.

\subsubsection{UAV networking}
Unmanned aerial vehicles (UAVs) often operate in dynamic environments and need to process high volumes of sensory data for navigation, obstacle avoidance, and mission management. SSL can enable UAVs to learn from vast amounts of unlabelled aerial data and improve over time without human supervision.
SSL can help in creating robust models that interpret environmental data, enhancing autonomous decision-making capabilities under varying conditions.

\subsubsection{Intelligent Reflecting Surface (IRS)}
IRS present a promising application scenario for SSL due to their ability to dynamically control and optimize wireless environments without direct human intervention. 
IRSs can manipulate electromagnetic waves to improve signal quality, coverage, and energy efficiency by reflecting signals in desired directions. SSL can be harnessed to train IRS systems to autonomously learn and adapt to varying environmental conditions and user demands. This learning paradigm enables the IRS to optimize its reflection parameters based on the observed performance metrics and the received signal patterns, enhancing the network performance without requiring labeled data.

\subsubsection{Network Security}

Network security represents a compelling application scenario for SSL due to its ability to autonomously identify patterns and anomalies within vast datasets without requiring labeled data. 
This learning paradigm is particularly advantageous in cybersecurity, where new types of threats emerge continuously and labeled examples of these threats are often unavailable or insufficient. By leveraging SSL techniques, network security systems can enhance their ability to detect previously unseen attack vectors, recognize unusual patterns indicative of potential breaches, and adapt to evolving threats in real-time. 

In summary, SSL has demonstrated benefits in various applications within wireless networks and holds tremendous potential when combined with emerging technologies.

\section{Case Study: Application of Self-Supervised Learning in Semantic Communication}


In this section, 
SSL techniques are used to learn semantic representations without extensive labeled datasets, offering a cost-effective and scalable solution.

\subsection{Implementation of SSL in Semantic Communication}

\begin{figure*}
    \centering
    \includegraphics[width=0.8\linewidth]{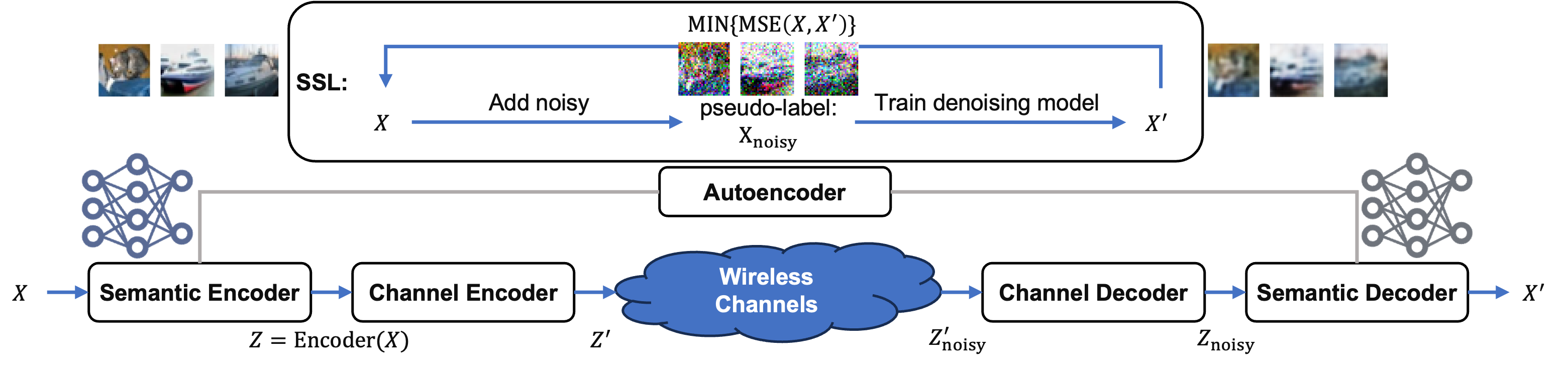}
    \caption{SSL enabled semantic communication system model: The autoencoder is trained to extract semantic features and reconstruct the original input. This is achieved through a self-supervised training process where the model learns to denoise the input data.}
    \label{fig_semcomm_ssl_model}
\end{figure*}

In our case study, we explore the application of SSL in a semantic communication system designed for image transmission over a wireless network as shown in Fig. \ref{fig_semcomm_ssl_model}. The primary goal is to encode and decode images in a manner that preserves the semantic content, even in the presence of noise.

\textbf{Data Preparation and Preprocessing:} 
We utilized the CIFAR-10 dataset, which includes $60,000$ $32\times 32$ color images in $10$ classes, with $50,000$ for training and $10,000$ for testing. The images were normalized to a mean of $(0.5, 0.5, 0.5)$ and a standard deviation of $(0.5, 0.5, 0.5)$.
The training set was loaded in batches of $128$ images.


\textbf{Model Architecture:}
We implemented an autoencoder for SSL model in semantic communication which consisted of a series of convolutional layers for the encoder and transposed convolutional layers for the decoder. 


\textbf{Training Process:}
To ensure reproducibility, we set random seeds for both NumPy and PyTorch. The models were trained using the Adam optimizer with a learning rate of $0.001$. The training of each model spanned $20$ epochs.
The self-supervised autoencoder was trained using mean squared error loss. 
Gaussian noise with a factor of $0.5$ was added to the input images to create noisy images, which the autoencoder then tried to reconstruct.


\subsection{Simulation Results}

An Ubuntu 20.04-based experimental platform powered by an Intel Xeon Gold 5218R CPU and an NVIDIA GeForce RTX 3090 GPU is utilized for the case study.
The performance of the model is evaluated by the Peak Signal-to-Noise Ratio (PSNR) metric.



\begin{figure}[!t]
\centering
\includegraphics[width=0.8\linewidth]{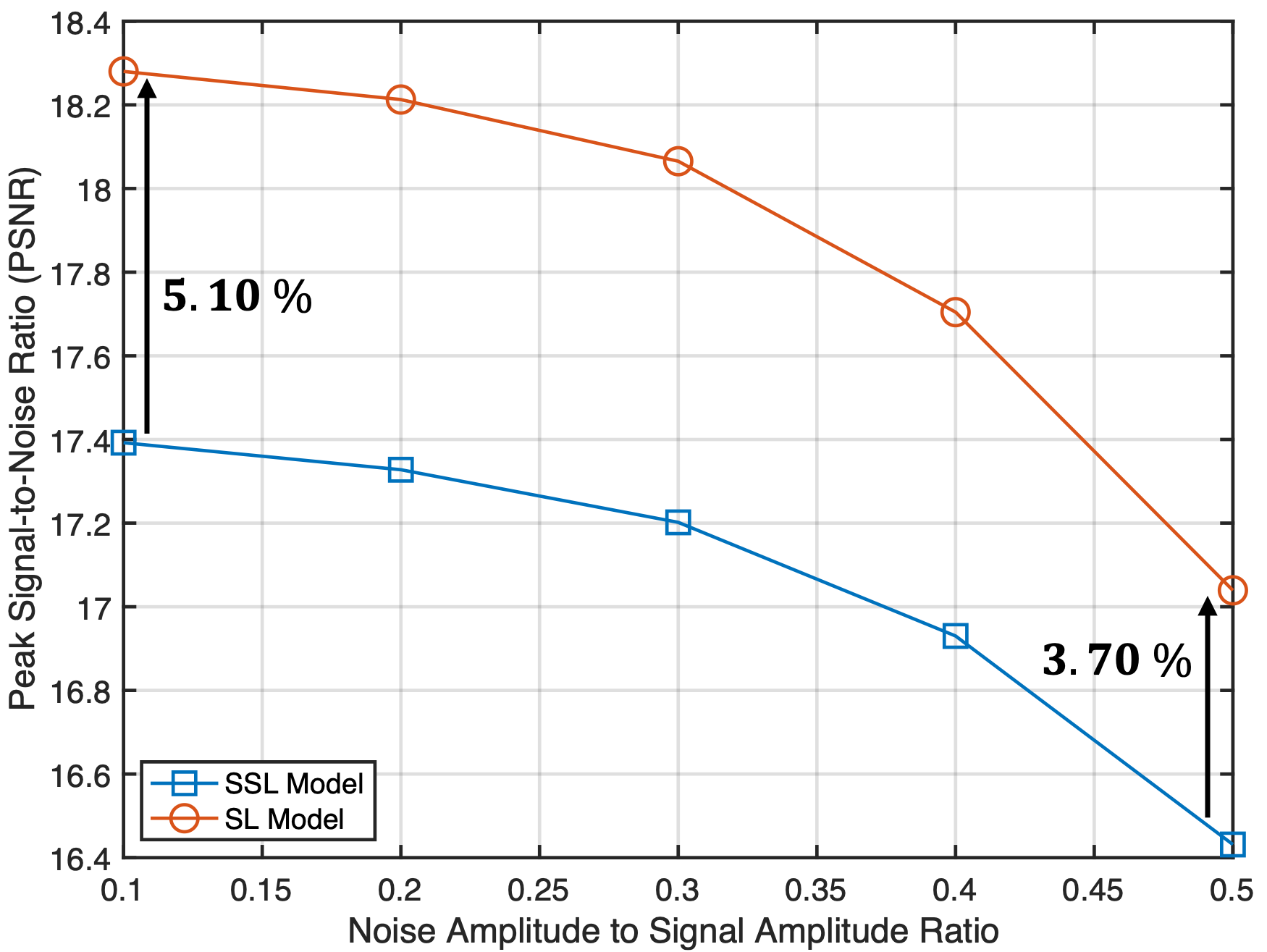}
\caption{PSNR versus NASAR.}
\label{fig:psnr vs snr}
\end{figure}

\begin{figure}[!t]
\centering
\includegraphics[width=0.8\linewidth]{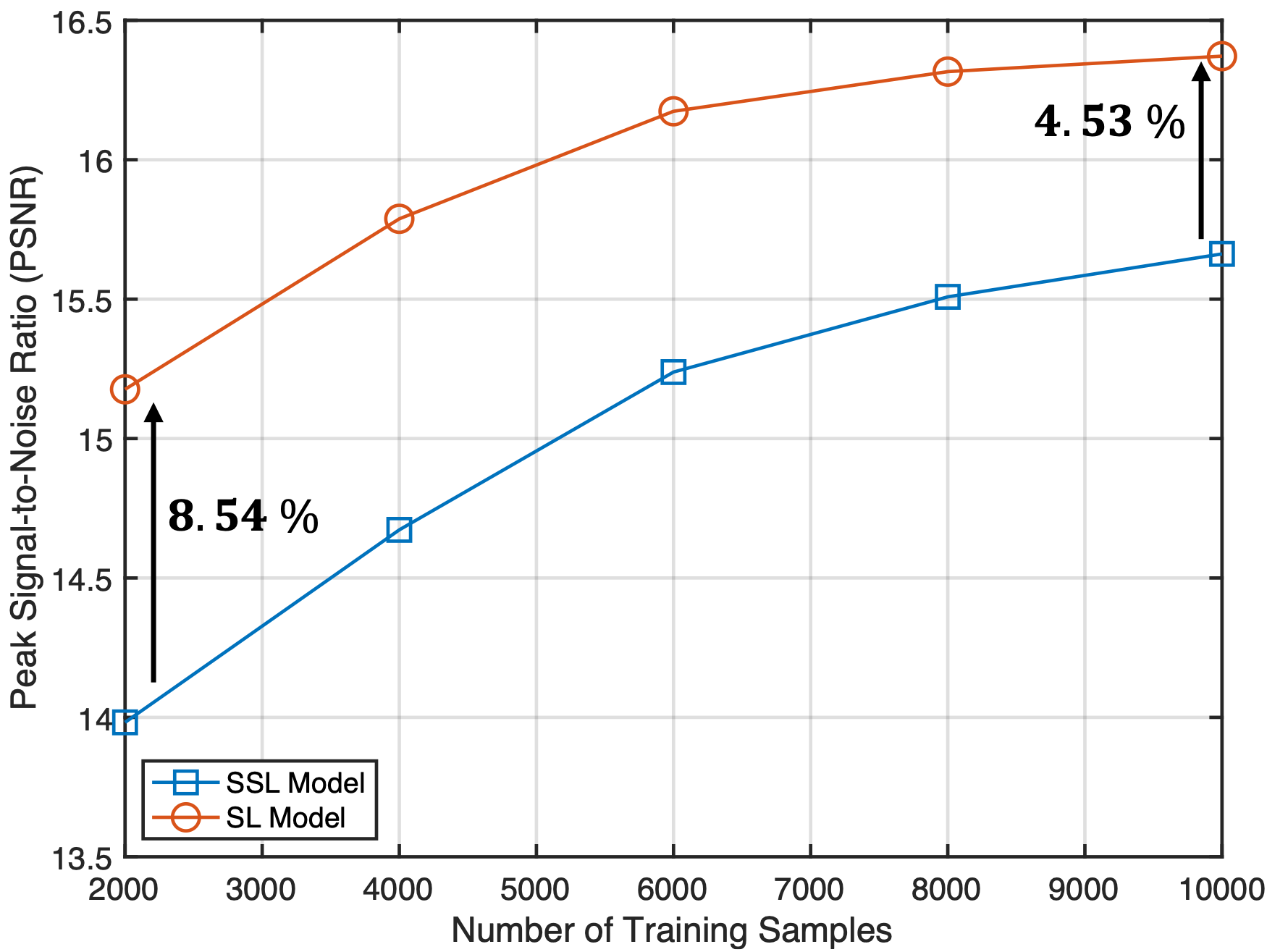}
\caption{PSNR versus number of training samples.}
\label{fig:psnr vs samples}
\end{figure}
Fig. \ref{fig:psnr vs snr} illustrates the PSNR of our proposed SSL model and supervised learning (SL) model under different noise amplitude to signal amplitude ratio (NASAR).
Overall, the SL model demonstrated superior performance in terms of PSNR across all NASAR values. Both models exhibited a decline in PSNR as the NASAR increased. 
However, the SSL model showed a relatively stable PSNR performance.
Specifically, when the noise amplitude ratio is $0.1$, the gap between the proposed model and the SL model is $5.10~\%$. When the noise amplitude increases to $0.5$, the gap between the proposed model and the SL model is further reduced to $3.70~\%$.
Fig. \ref{fig:psnr vs samples} illustrates the PSNR of our proposed SSL model and SL model under different number of training samples.
The SL model outperformed the SSL model in terms of PSNR across all sample sizes. Both models showed improved PSNR with an increasing number of training samples. 
However, the performance gain of SSL model (gap reduced from $8.54~\%$ to $4.53~\%$) with more samples underscores its efficiency and effectiveness in leveraging large datasets without the necessity for labeled data, thus offering significant advantages in resource-constrained environments.

While the SL model consistently provided higher PSNR values compared to the SSL model across different SNR levels and training sample sizes, the SSL approach demonstrated notable strengths. 
Specifically, the SSL model achieved competitive PSNR performance without requiring labeled data, making it a practical and efficient alternative in scenarios with limited labeled data availability.
Additionally, the stable performance of SSL model across varying conditions emphasizes its robustness and adaptability, suggesting a viable and scalable solution for real-world applications where data labeling is a bottleneck.


\section{FUTURE DIRECTIONS}

There are several open challenges in using SSL in the wireless network optimization. 
We elaborate on several of them in this section.



\subsection{Dynamic and Evolving Environments}
Wireless networks are dynamic and continuously evolving due to changes in user behavior, device mobility, and environmental conditions. SSL models must adapt to these changes to remain effective.
Future research should focus on developing adaptive SSL frameworks that can quickly respond to network changes. Techniques such as online learning, continual learning, and transfer learning could be leveraged to maintain model relevance over time.

\subsection{Interpretability and Explainability}
SSL models can be complex and opaque, making it difficult for network operators to understand and trust their decisions. Interpretability and explainability are crucial for the practical deployment of SSL in wireless networks.
Investigating methods to enhance the transparency of SSL models, such as interpretable machine learning techniques and explainable AI, will be important. Providing clear and actionable insights derived from SSL models can facilitate their adoption in real-world scenarios.


\subsection{Security and Privacy Concerns}
SSL in wireless networks may raise security and privacy concerns, especially when dealing with sensitive user data. Ensuring the confidentiality and integrity of data used for learning is paramount.
Addressing security and privacy issues through techniques such as federated learning, differential privacy, and secure multiparty computation can help mitigate risks. Research should aim to balance optimization performance with stringent security and privacy requirements.



\section{CONCLUSION}

In this article, we have provided a comprehensive
overview of SSL in wireless network optimization.
The case study illustrates the potential of SSL to enhance semantic communication in wireless networks, especially in scenarios where noise interference is strong and data is difficult to obtain in large quantities.
The PSNR gap between SSL model and SL model can be narrowed to less than $5~\%$.
However, open research issues remain, such as interpretability and privacy concerns.
Overall, this paper provides valuable insights into the potential of SSL in wireless network.
As wireless networks continue to evolve, integrating SSL techniques will be crucial in addressing the growing demand for efficient and reliable communication.


\bibliographystyle{IEEEtran}
\bibliography{references}


\end{document}